\documentclass[12pt]{article}

\usepackage{url}

\usepackage{scicite}

\usepackage{times}

\newenvironment{sciabstract}{%
\begin{quote} \bf}
{\end{quote}}

\newcounter{lastnote}
\newenvironment{scilastnote}{%
\begin{list}%
{\arabic{lastnote}.}
{\usecounter{lastnote} 
\setcounter{lastnote}{\value{enumiv}}
\setlength{\leftmargin}{.26in}}
{\setlength{\labelsep}{.5em}}
}
{\end{list}}

\usepackage[dvips]{graphicx}
\usepackage{amssymb,amsfonts,amsmath}
\usepackage{array}
\usepackage{dcolumn}
\usepackage{booktabs}

\usepackage{amstext}

\newcommand{\hpp}{homogeneous Poisson process}
\newcommand{\cpp}{cascading Poisson process}
\newcommand{\nscpp}{non-stationary cascading Poisson process}
\newcommand{\nhpp}{non-homogeneous Poisson process}

\newcommand{\cnhpp}{cascading non-homogeneous Poisson process}
\newcommand{\pvalue}{$p$-value}

\newcommand{\suppinfo}[1]{#1}
\newcommand{\figref}[1]{Fig.~#1}
\newcommand{\figsref}[1]{Figs.~#1}

\newcommand{\secref}[1]{Sec.~#1}

\newcommand{\tblref}[1]{Tbl.~#1}
\newcommand{\tblsref}[1]{Tbls.~#1}

\newcommand{\dataparsingsec}{S1}
\newcommand{\robustsec}{S2}
\newcommand{\othermodelssec}{S3}
\newcommand{\cppanalsec}{S4}
\newcommand{\mchtsec}{S5}

\newcommand{\allSItbls}{S1--S3}

\newcommand{\cppparamfig}{S4}
\newcommand{\cppptaufig}{S5}
\newcommand{\allSIfigs}{S1--S5}

\newcommand{\nptsave}{100}

\newcommand{\nindividuals}{16}
\newcommand{\occupations}{writers, performers, politicians, and scientists}

\newcommand{\nptspersegment}{10}
\newcommand{\pvalcutoffpercent}{5\%}
\newcommand{\pvalcutoffdecimal}{0.05}

\newcommand{\neinsteinsegments}{54}

\newcommand{\iii}{i}  

\newcommand{\rate}{\rho}

\newcommand{\bftheta}{\boldsymbol{\theta}}
\newcommand{\bfthetai}{\bftheta_{\iii}}

\newcommand{\ratei}{\rho_{\iii}}
\newcommand{\qi}{\xi_{\iii}}

\newcommand{\prob}[1]{p\left(#1\right)}

\newcommand{\cdf}[1]{P(#1)}

\newcommand{\panela}{A}
\newcommand{\panelb}{B}
\newcommand{\panelc}{C}
\newcommand{\paneld}{D}
\newcommand{\panele}{E}

\begin{document}
\baselineskip24pt

\title{On universality in human correspondence activity
}

\author{%
R.~Dean Malmgren$^{1\ast}$,
Daniel B.~Stouffer$^{1,2}$, \\
Andriana S.~L.~O.~Campanharo$^{1,3}$,
Lu\'is A.~Nunes Amaral$^{1,4\ast}$\\
\\
\normalsize{$^1$Department of Chemical and Biological Engineering, Northwestern
  University,}\\ \normalsize{Evanston, IL 60208, USA} \\
\normalsize{$^2$Integrative Ecology Group, Estaci\'on Biol\'ogica de Do\~nana, CSIC, 41092 Sevilla,
  Spain} \\
\normalsize{$^3$Instituto Nacional de Pesquisas Espaciais, 12227-010 S\~ao
  Jos\'e dos Campos,}\\ \normalsize{S\~ao Paulo, Brazil} \\
\normalsize{$^4$Northwestern Institute on Complex Systems, Northwestern
  University,}\\ \normalsize{Evanston, IL 60208, USA} \\
\\
\normalsize{$^{\ast}$To whom correspondence should be addressed;}\\ 
\normalsize{E-mail: dean.malmgren@u.northwestern.edu, amaral@northwestern.edu.}
}

\date{}

\maketitle

\begin{sciabstract}
Identifying and modeling patterns of human activity has important ramifications
in applications ranging from predicting disease spread to optimizing resource
allocation.  Because of its relevance and availability, written correspondence
provides a powerful proxy for studying human activity.  One school of thought
is that human correspondence is driven by responses to received correspondence,
a view that requires distinct response mechanism to explain e-mail and letter
correspondence observations.  Here, we demonstrate that, like e-mail
correspondence, the letter correspondence patterns of
\nindividuals~\occupations~are well-described by the circadian cycle, task
repetition and changing communication needs.  We confirm the universality of
these mechanisms by properly rescaling letter and e-mail correspondence
statistics to reveal their underlying similarity.  

\end{sciabstract}

\renewcommand{\baselinestretch}{2.0}

Power-law statistics are a hallmark of critical phenomena.  A less obvious
characteristic of criticality is the emergence of universality classes that
capture the similarity of seemingly disparate systems.  For example, despite
the fact that water and carbon dioxide have different chemical properties, they
were observed to behave in the same manner close to their respective critical
points~\cite{stanley71}.  This is because idiosyncrasies, such as the existence
of electric dipoles or the ability to form hydrogen bonds, become irrelevant
near the liquid--gas critical point.
For physical systems, renormalization group theory~\cite{ma76,goldenfeld92} has
enabled researchers to understand the deep connection between the symmetries of
a system and the mechanisms which underlie its behavior.  The similarity of
different fluids near their respective liquid--gas critical points is often
demonstrated by rescaling their statistics such that they collapse onto the
same \emph{universal} curves---oftentimes power-laws which have particular
scaling exponents.
By grouping different substances into the same ``universality class,'' as
identified by its scaling exponents, one uncovers that fluids are described by
the same statistical laws near the liquid-gas critical point as uniaxial
magnets are near their paramagnetic critical point~\cite{stanley71}.
Importantly, one can also differentiate the behavior of these systems from
the behavior of polymers near the sol-gel transition, which belong to a
different universality class~\cite{stanley71}.

In addition to critical phenomena, power-law scaling has also been widely
reported in biology, economics, and
sociology~\cite{bachelier00,pareto06,auerbach13,gibrat31,zipf49,ijiri77,newman05a}.
Renormalization group theory therefore offers a tantalizing hypothesis for the
prevalence of particular power-law scaling exponents in social systems: social
systems, in analogy with physical systems, may operate near critical points and
can therefore be classified into a small number of distinct universality
classes.
A heated debate has consequently ensued in the literature concerning the
``universality of human systems'' (in the statistical physics meaning of the
phrase).  Is there enough statistical evidence for the asymptotic power-law
description of the heavy-tailed distributions reported in human
systems~\cite{avnir98,amaral00,clauset07,malmgren08}?  Is it reasonable to
postulate that social systems, like their physical
counterparts~\cite{ma76,goldenfeld92,stanley00}, can be classified into
universality classes according to scaling exponents~\cite{stanley96a}?

Human correspondence is a paradigmatic area where the matter of power-law
scaling and universality are contentious issues.  One view that has recently
received significant attention in the literature~\cite{castellano07,zhou08}
posits that correspondence patterns are driven primarily by the need to respond
to other individuals.  This is formalized by a priority queuing
model~\cite{barabasi05} which, under certain limiting conditions, reproduces
the asymptotic scaling of empirically observed heavy-tailed correspondence
statistics.
In particular, the heavy-tailed statistical properties of e-mail correspondence
are reportedly reproduced by a fixed-length queue with a single task
type~\cite{barabasi05,vazquez06} whereas the heavy-tailed statistical
properties of letter correspondence are reportedly reproduced by either a
variable-length queue with a single task type~\cite{oliveira05,vazquez06} or by
a fixed-length queue with multiple task types~\cite{grinstein06}.
The fact that there are different exponents for the two modes of correspondence
has been taken as evidence that human correspondence falls into one of two
universality classes~\cite{vazquez06}.  When interpreted in the statistical
mechanics sense of ``universality,'' one would conclude that e-mail and letter
correspondence are \emph{fundamentally} different activities.

In contrast, we hypothesize that human correspondence patterns are not driven
by responses to others but by more prosaic mechanisms---circadian cycles, task
repetition and changing communication needs.  We formalize these mechanisms
with a \cnhpp, which we have previously shown to be statistically consistent
with e-mail communication patterns~\cite{malmgren08}.  Here, we hypothesize
that the same model is capable of describing letter correspondence and that the
heavy-tailed correspondence statistics primarily arises from the variation in
an individual's communication needs over the course of their lifetime.

We obtained the letter correspondence records for \nindividuals~\occupations.
Each data set consists of a list of letters that were sent by each of these
individuals, and each record comprises the name of the sender, the name of the
recipient, and the date when it was written
(see \suppinfo{\secref{\dataparsingsec}~for details}).
The nature of the data raises two issues to consider during analysis.  
First, the precise authorship date of some letters is unknown, so we restrict
our analysis to only those letters that have \emph{precise} authorship dates.
Second, it is highly unlikely that all of the letters written by a particular
individual are present in the database.  We have confirmed that our
results are insensitive to sampling effects from this method of data collection
(\suppinfo{\secref{\robustsec}}).

\begin{figure}
\centerline{\includegraphics*[height=0.6\textheight]{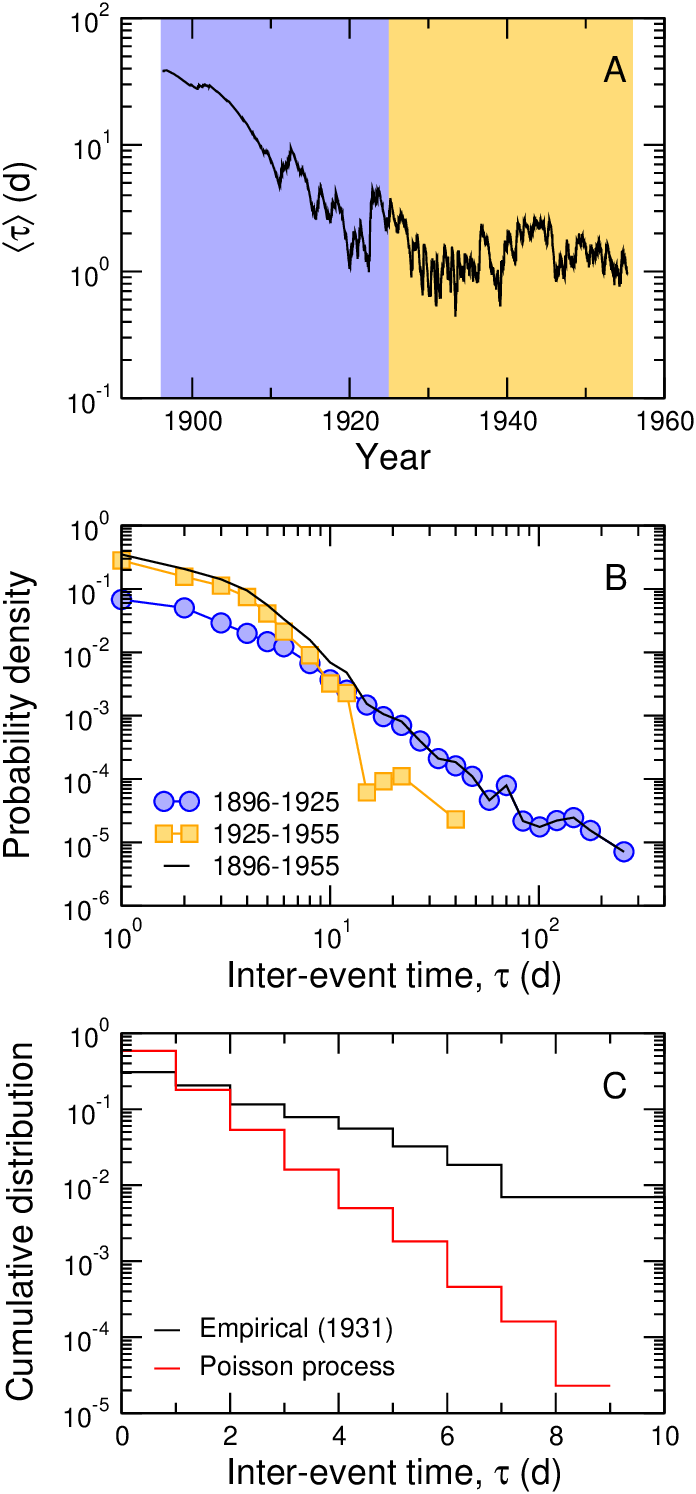}}
\caption{~
}
\label{fig:running_ave}
\end{figure}

An important consideration in studying the letter correspondence
patterns of these individuals is that the data covers their
\textit{entire} lifetimes.  As a result, it is quite conceivable that
changing communication needs might affect letter correspondence patterns.
For example, before Einstein became widely known, the bulk of his recorded
communication was to friends and relatives.  After the confirmation of his
theory of relativity in 1919, Einstein's need to communicate with other
individuals substantially increased.  By that time, his step-daughter Ilse
Einstein was helping him with secretarial tasks resulting in greatly improved
coverage of his recorded correspondence~\cite{sauer09}.
Due to this secretarial assistance and his increased fame, we expect that the
average time between consecutively sent letters, the average inter-event time
$\langle\tau\rangle$, is significantly larger during the beginning of
Einstein's life than during the latter part of his life.  Our expectations are
verified in
\figref{\ref{fig:running_ave}\panela--\panelb}, demonstrating that these time
series' are non-stationary---that is, the heavy-tailed inter-event time
distribution results from a mixture of time scales~\cite{hausdorff96}.

Since these time series' are non-stationary, we partitioned each complete time
series into smaller time segments so that we can make the approximation that
the behavior within each time segment is stationary.  
%
%
We accomplish this by splitting the time series into segments lasting
364 days (52 weeks) unless fewer than \nptspersegment~events fall within that
time period, in which case consecutive segments are merged until this criterion
is met.

Assuming that the correspondence patterns within each time segment are
stationary, we can then model the behavior within each time segment with
standard techniques.  As a first approximation, one might na\"ively expect that
letters are sent at a constant rate $\rate$ and that the time at which every
letter is sent is independent of all others.  Such a process is referred to as
a \hpp, which gives rise to an exponential inter-event time distribution
$\prob{\tau}=\rate e^{-\rate\tau}$.  While the tail of the inter-event time
distribution within these time segments is approximately exponential, the
best-estimate predictions of a \hpp~does not produce the correct decay rate
(\figref{\ref{fig:running_ave}\panelc}).  This suggests that only a few changes
to the \hpp~are needed to statistically reproduce the observed inter-event time
distribution.  We hypothesize that, like e-mail correspondence, two additional
ingredients must also be considered for modeling letter
correspondence~\cite{malmgren08}.

First, daily and weekly cycles of activity may influence when individuals
communicate.  Previously, we accounted for these cycles of activity in e-mail
communication with a \nhpp~whose rate $\rate(t)$ changes periodically on 
daily and weekly time scales.  For letter correspondence, however, the
resolution of the data does not permit us to identify activity patterns within
a day, and day-to-day changes in activity provide no additional insight
(\suppinfo{\secref{\othermodelssec}}).  We therefore approximate
the \nhpp~defined by $\rate(t)$ by a \hpp~with constant rate $\ratei$ during
time segment $i$; that is, we model the rate of activity $\rate(t)$
throughout each individual's life by a piecewise constant function of time.

Second, individuals are much more likely to continue writing letters once they
have written one letter in order to use their time more effectively.  We
account for this behavior by hypothesizing that, once an individual finishes
writing a letter, there is a probability $\qi$ that they write another letter.
This process repeats itself until this \emph{cascade} of additional letters
concludes with probability $1-\qi$, at which point the individual's behavior is
again governed by a \hpp~with rate $\ratei$~\cite{footnote1}.  We refer to the
resulting model as a \emph{\cpp}.


To compare the predictions of the \cpp~\cite{sciencemethods09} to the empirical
data, we must first estimate the parameters $\bfthetai=\{\ratei,\qi\}$ from the
data during each time segment.  The nature of the data, however, raises an
important concern for parameter estimation: since each event is only known to
occur within a particular day, not at a precise time of the day, the data are
interval censored~\cite{dagostino86}.
We account for the interval censored data and calculate the best-estimate
parameters $\widehat{\bftheta}_{\iii}$ by numerically maximizing the censored
likelihood function (see \secref{\cppanalsec}~for the derivation).

The resulting best-estimate parameters $\widehat{\bftheta}_{\iii}$ provide
insight into the correspondence patterns of each individual
(\figref{\ref{fig:fig2}\panela--\panelb}~and \figref{\cppparamfig}).
For example, while both Schoenberg and Einstein have a 50-fold increase in the
rate at which the send letters---presumably due to their increasing
correspondence obligations and a more complete sampling of their overall letter
correspondence---their utilization of cascades of activity is markedly
different.  Schoenberg, for instance, sends about 21\% of his letters during
cascades of multiple letters throughout his life.  In contrast, Einstein rarely
utilizes cascades of activity as a young man (before 1910) whereas in later
years (after 1933) he sends approximately 34\% of his letters during cascades
of multiple letters.

In the period 1928--1933, Einstein sent over 50\% of his letters
during cascades of multiple letters.  The start of this period
coincides with the hiring of Einstein's long-time secretary Helen Dukas, who
more systematically retained copies of his outgoing correspondence.  After the
Nazis took over power in January 1933, his correspondence patterns change
markedly; this possibly reflects changes in his correspondence obligations at
Princeton University after immigrating to the United States in late
1933~\cite{sauer09}.

Of course, inferring how an individual's behavior changes based on a model's
parameter estimates is contingent upon the model being consistent with the data.
We tested the statistical consistency of our model with the data by Monte
Carlo hypothesis testing (\suppinfo{\secref{\mchtsec}}).  We reject
the model during a particular time segment if the \pvalue~obtained from the
Monte Carlo hypothesis testing procedure is less than a threshold
of \pvalcutoffdecimal.  Because this threshold is greater than zero, it means
that there is a finite chance that we will reject the hypothesis that the model
is consistent with the data even if the data was generated from the model.

If we assume that each time segment is independent, then we would expect to
reject each of the time segments with a 
\pvalcutoffpercent~chance and the total number of rejections to
be distributed according to a binomial model~\cite{miller91}.  Out of
the \neinsteinsegments~independent time segments for Einstein for example, we
would expect to reject the model 2.7 times with 0--6 defining the bounds of the
95\% confidence interval of the corresponding binomial model.  For Einstein,
our procedure ``rejects'' the \cpp~for 2 out of 54 time segments, indicating
that we cannot reject the hypothesis that the model is able to explain his
correspondence patterns.  Indeed, our hypothesis testing confirms that
the \cpp~can not be rejected as an explanatory model for the letter
correspondence of \emph{any} of the individuals under consideration
(\tblref{\ref{tbl:summary}}).  These results demonstrate that the origin of the
heavy-tailed inter-event time distribution is a mixture of distributions with
different time scales (\figref{\ref{fig:fig2}\panelc--\panele}).

Our findings enable us to address a crucial question: do e-mail and letter
correspondence belong to different universality classes~\cite{vazquez06}?
Since the same mechanistic model is capable of describing both e-mail and
letter correspondence, we can answer this question in the negative.  We
demonstrate the underlying similarity of both correspondence activities by
rescaling and collapsing the inter-event time distributions
for \nindividuals~randomly selected e-mail correspondents~\cite{eckmann04} for
which we have model parameter estimates~\cite{malmgren08} and
the \nindividuals~letter correspondents studied here
(\figref{\ref{fig:rescaled_tau}}).  The rescaled inter-event time distributions
agree with theoretical expectations~\cite{taylor94}, demonstrating that the
same exponential statistical law is indeed capable of describing both
correspondence patterns.

Only by understanding and validating the underlying mechanisms can we
appropriately rescale e-mail and letter correspondence to reveal their
underlying similarity.  Unlike critical phenomena, the universality here does
not arise from the irrelevance of idiosyncrasies but rather from the fact that
these two different modes of communication are governed by the same mechanisms.
This insight is not apparent just by studying the asymptotic scaling of an
empirical distribution obtained from an individual; one simply cannot infer
that different ``scaling'' exponents necessarily imply different mechanisms.

Our results therefore raise significant questions about the nature of
universality in complex phenomena, in general, and in human correspondence, in
particular.  Perhaps the most common universal statistical law is due to the
central limit theorem---sums of variates with finite fluctuations converge to a
Gaussian distribution.  When confronted with statistical patterns that are
non-Gaussian one is tempted to surmise that the system's fluctuations are not
finite.  In analogy to physical systems, the recurrence of power-law
dependencies with similar exponent values in biological or social systems is
frequently hypothesized to arise from the fact that these systems operate near
critical points where particular details of the system become irrelevant.

A less-explored hypothesis, as exemplified here, is that heavy-tailed
distributions emerge as a result of non-stationarities in the absence of
criticality~\cite{malmgren08,amaral98}.  Our study demonstrates that human
correspondence can be accurately modeled as a \cnhpp---a non-critical process.
This process gives rise to heavy-tailed statistics but not to power-law
statistics characterized by critical exponents.  Instead, the correspondence
patterns of each individual are uniquely characterized by the parameters of our
model~\cite{malmgren09a}; the process is universal, but the parameters are not.

Indeed, we postulate that the cascading Poisson process, which formally
incorporates the circadian cycle, task repetition and changing needs, may
accurately describe many other aspects of human activity.  The circadian cycle
has such physiologic impact that it is natural to surmise that it will affect
numerous human behaviors, from eating habits to commuting routines.  Task
repetition is similarly important due to the increased efficiency it enables;
once an individual makes one purchase at a mall, it is easier to make other
purchases within that mall during the same trip than it is to return to the
mall the following day.  As one ages and changes roles, it is not hard to
imagine that the extent with which the circadian cycle and task repetition
influence their activity might change over time.  It is therefore plausible
that the cascading Poisson processes presented here could be generalized to
account for different types of activities, each with its own evolving
parameters.



\vspace*{1cm}
\bibliographystyle{Science}

\begin{thebibliography}{10}
\bibitem{footnote1}
We assume that $\qi\gg\ratei dt$. This model formulation is equivalent to
  having $N_a$ additional events where $N_a$ is drawn from an exponential
  distribution with mean $\qi/(1-\qi)$.
\end{thebibliography}

\begin{scilastnote}
\item We thank T.~Sauer, R.~Guimer\`a, M.~Sales-Pardo, M.J.~Stringer,
E.N.~Sawardecker, J.~Duch and P.~McMullen for insightful comments and
suggestions.
D.B.S.~acknowledges the support of a CSIC-JAE Postdoctoral Fellowship.
A.S.L.O.C.~acknowledges the support of a CNPq (Brazil) Doctoral Fellowship.
L.A.N.A.~gratefully acknowledges the support of NSF award SBE 0624318.
All figures were generated with PyGrace (http://pygrace.sourceforge.net) with
color schemes from http://colorbrewer.org.

\item 
\textbf{Supporting Online Material} \\
www.sciencemag.org \\
Materials and Methods \\
Supporting text \\
\figsref{\allSIfigs} \\
\tblsref{\allSItbls}

\end{scilastnote}


\clearpage

\begin{table}
\small
\centering
\caption{
Summary of the letter correspondence records and hypothesis testing results for
the \nindividuals~individuals under consideration, ordered chronologically.
For each individual, we note
the time period and duration of the letter correspondence records,
the total number of letters sent, 
the number of time segments with at least \nptspersegment~letters per time
segment,
the 95\% confidence interval (CI) bounds of the corresponding binomial model
with $p=\pvalcutoffdecimal$,
and the number of rejections of the \cpp~based on our Monte Carlo hypothesis testing
procedure.
The number of Monte Carlo hypothesis testing rejections is within the 95\%
confidence interval bounds for all \nindividuals~individuals, indicating that
this model can not be rejected for \emph{any} individual's letter
correspondence patterns.  We have conducted the same analysis for three
alternative models; we find that a \cpp~provides the most parsimonious and
statistically consistent explanation of the data
(\suppinfo{\secref{\othermodelssec}}).
}
\medskip
\label{tbl:summary}
\renewcommand{\baselinestretch}{1.0}
\begin{tabular}{lr@{--}lcD{.}{.}{6.0}D{.}{.}{2.0}c*{1}{D{.}{.}{3.1}}}
\toprule
           & \multicolumn{2}{c}{Time}   & Duration &
\multicolumn{1}{c}{Number of} & \multicolumn{1}{c}{Number of} & & 
\multicolumn{1}{c}{Number of} \\ [-2ex]
Individual & \multicolumn{2}{c}{Period} & (yr) &
\multicolumn{1}{c}{letters} & \multicolumn{1}{c}{segments} & 95\% CI & 
\multicolumn{1}{c}{rejections} \\
\midrule 
Francis Bacon & 1574 & 1626 & 53 & 443 & 19 & $\left[0,3\right]$ & 3 \\ [-2ex]
James H.~Leigh Hunt & 1790 & 1859 & 70 & 408 & 25 & $\left[0,3\right]$ & 1 \\ [-2ex]
Charles Darwin & 1822 & 1882 & 61 & 6,785 & 52 & $\left[0,5\right]$ & 4 \\ [-2ex]
Anna Brownell Jameson & 1833 & 1860 & 28 & 119 & 8 & $\left[0,2\right]$ & 1 \\ [-2ex]
Friedrich Engels & 1833 & 1895 & 63 & 369 & 24 & $\left[0,3\right]$ & 1 \\ [-2ex]
Robert E.~Lee & 1835 & 1870 & 36 & 282 & 10 & $\left[0,2\right]$ & 0 \\ [-2ex]
Karl Marx & 1837 & 1882 & 46 & 469 & 25 & $\left[0,3\right]$ & 1 \\ [-2ex]
Henry Irving & 1852 & 1905 & 54 & 1,205 & 35 & $\left[0,4\right]$ & 0 \\ [-2ex]
Sigmund Freud & 1872 & 1939 & 68 & 3,130 & 49 & $\left[0,5\right]$ & 2 \\ [-2ex]
Marcel Proust & 1879 & 1922 & 44 & 668 & 25 & $\left[0,3\right]$ & 2 \\ [-2ex]
H.~G.~Wells & 1895 & 1946 & 52 & 422 & 16 & $\left[0,2\right]$ & 0 \\ [-2ex]
Albert Einstein & 1896 & 1955 & 60 & 10,319 & 54 & $\left[0,6\right]$ & 2 \\ [-2ex]
Carl Sandburg & 1898 & 1966 & 69 & 1,894 & 37 & $\left[0,4\right]$ & 2 \\ [-2ex]
Arnold Schoenberg & 1902 & 1951 & 50 & 6,899 & 47 & $\left[0,5\right]$ & 3 \\ [-2ex]
Ernest Hemingway & 1909 & 1961 & 53 & 1,934 & 42 & $\left[0,5\right]$ & 5 \\ [-2ex]
Stan Laurel & 1924 & 1964 & 41 & 685 & 17 & $\left[0,3\right]$ & 1 \\
\bottomrule
\end{tabular}
\end{table}

\clearpage

\clearpage

\noindent \figref{\ref{fig:running_ave}}:
Non-stationarity of Albert Einstein's letter correspondence activity.  While we
select Einstein as an example, non-stationarities are present for all
\nindividuals~\occupations~studied here.
{\bf \panela}, Running average inter-event time $\langle\tau\rangle$ averaged
over \nptsave~consecutive inter-event times.  During the beginning of
Einstein's life (blue shaded region), the average inter-event time is
significantly larger than during the end of his life (orange shaded region).
{\bf \panelb}, Logarithmically binned probability density of the
\textit{non-zero} inter-event times $\tau$.  If we separately consider the
inter-event time distribution during each portion of Einstein's life, it is
clear that the complete inter-event time distribution (black line) is actually
a mixture of behaviors.  To emphasize the origins of the heavy-tailed
distribution, the probability densities of each portion of Einstein's life are
normalized such that their integrals are equal to the fraction of non-zero
inter-event times during that time.
{\bf \panelc}, Comparison of the empirical inter-event time distribution during
a particular time segment with the simulated predictions of the best-estimate
\hpp~that is interval censored in the same manner as the data.  It is visually
apparent that a \hpp~is not consistent with the empirical data, which is
confirmed by Monte Carlo hypothesis testing
(\suppinfo{\secref{\othermodelssec}}).
%


\clearpage


\begin{figure}
\centerline{\includegraphics*[width=1.0\columnwidth,height=1.0\textheight,keepaspectratio=true]{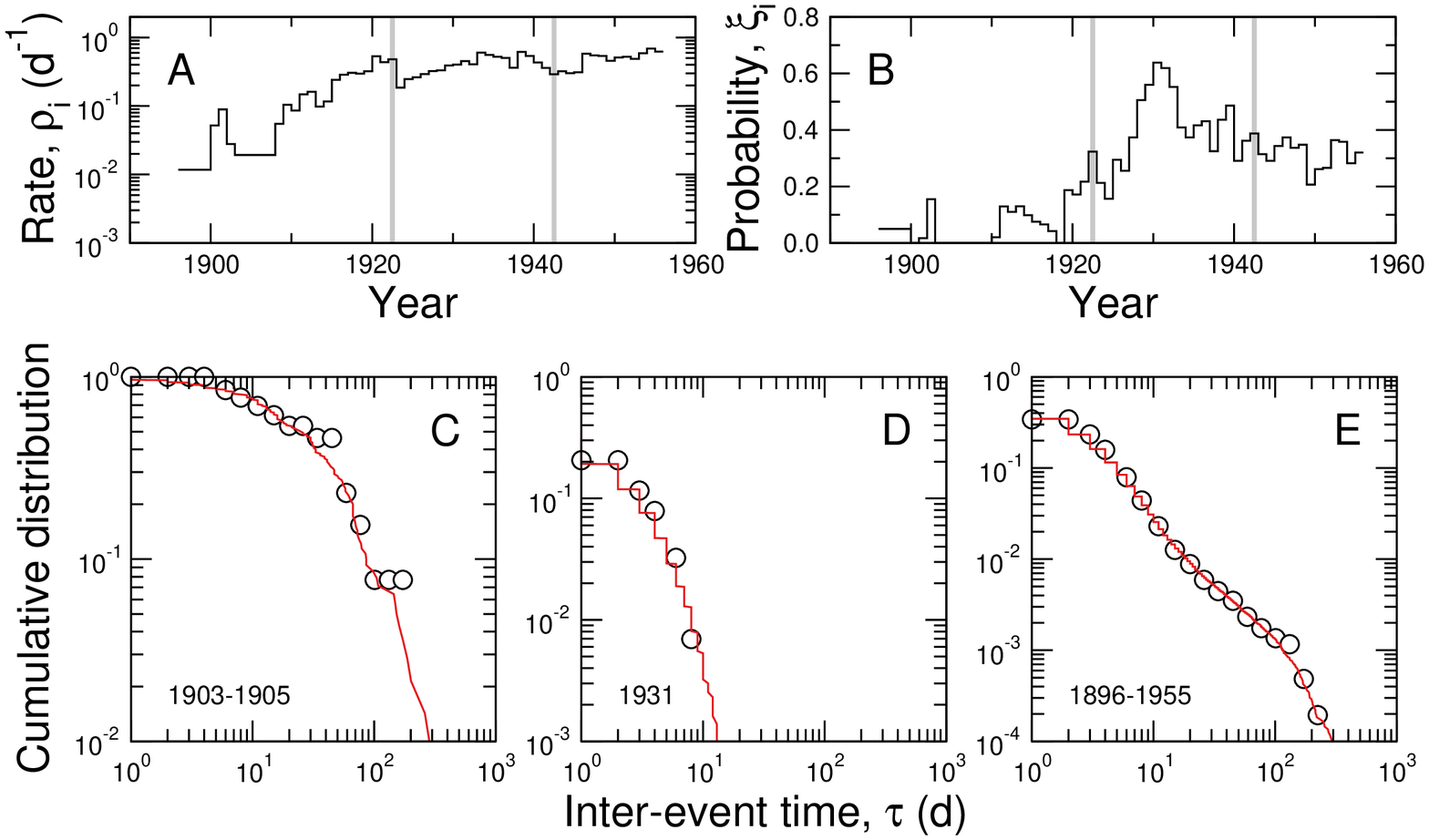}}
\renewcommand{\baselinestretch}{2.0}
\caption{~
}
\label{fig:fig2}
\end{figure}

\clearpage

\noindent \figref{\ref{fig:fig2}}:
Origin of heavy-tailed inter-event time distribution for Albert Einstein.
While we select Einstein as an example, the same explanation is relevant for
all \nindividuals~\occupations~studied here.
{\bf \panela}--{\bf \panelb}, We estimate the parameters
$\bfthetai=\{\ratei,\qi\}$ by maximizing the censored likelihood function for
each time segment (\suppinfo{\secref{\cppanalsec}}).  Grey shaded regions
denote time segments during which the \cpp~is rejected by Monte Carlo
hypothesis testing. Parameter estimates for all individuals under consideration
can be found in \figref{\cppparamfig}.  Note the 50-fold changes in the
rate $\ratei$ and the dramatic changes in $\qi$ for Einstein.
{\bf \panelc}--{\bf \paneld}, The cumulative distribution of inter-event times
for Einstein during particular time segments compared with the predictions of a
\nscpp~with the best-estimate parameters (\panela--\panelb).
The model predictions are generated numerically by running the model defined by
$\bftheta(t)$ ten-times and interval censoring the resulting synthetic time
series in the same manner as the empirical data.
{\bf \panelc}, The cumulative distribution of inter-event times (circles) for
Einstein over his entire life compared with the predictions of a \nscpp~(red
line) with the best-estimate parameters (\panela--\panelb).
When the mixture of behaviors are taken into account, the origin of the
heavy-tailed inter-event time distribution is clear.  The inter-event time
distributions for all \nindividuals~letter correspondents under consideration
can be found in \figref{\cppptaufig}.

\clearpage

\begin{figure}
\centerline{\includegraphics*[height=0.6\textheight]{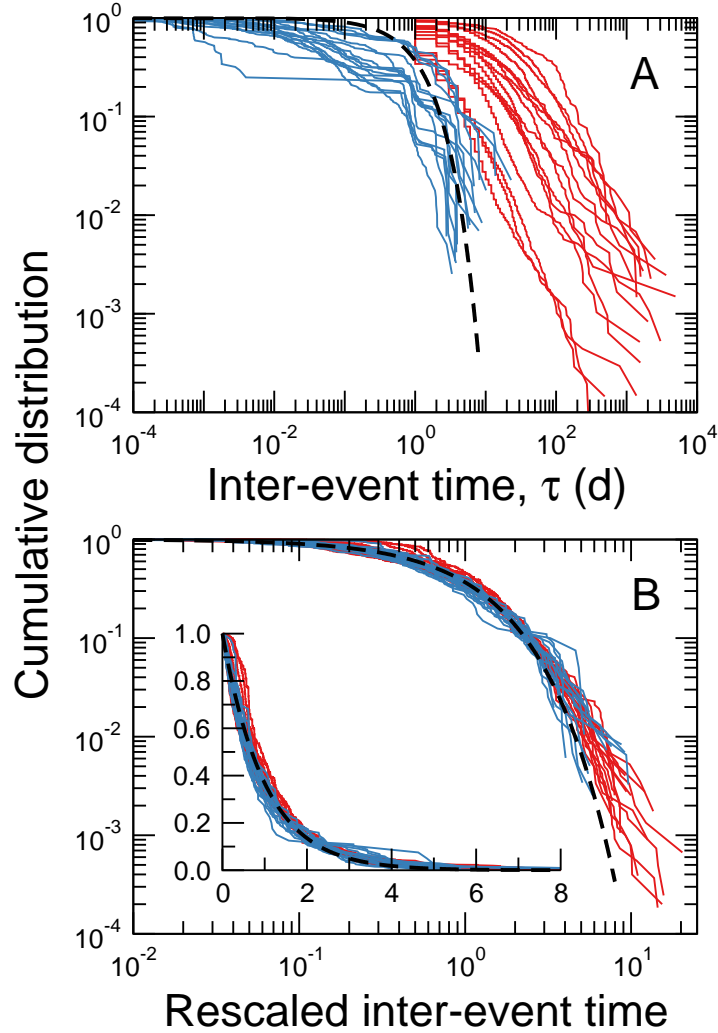}}
\caption{
Collapse of inter-event time distributions for letter and e-mail
correspondence.
{\bf \panela}, Cumulative distribution of inter-event times for all
\nindividuals~letter correspondents (red lines) and \nindividuals~randomly
selected e-mail correspondents (blue lines). 
{\bf \panelb}, Cumulative distribution of rescaled inter-event times on
logarithmic and linear (inset) axes.  The inter-event time
$\tau_k=t_{k+1}-t_{k}$ is rescaled by the average inter-event time expected
during the interval $\left[t_{k},t_{k+1}\right]$, which is given by
$\langle\tau\rangle=(t_{k+1}-t_{k})/\int_{t_{k}}^{t_{k+1}}\rho(s)ds$.  By the
time rescaling theorem~\cite{taylor94}, the resulting rescaled inter-event time
distribution is given by the expected inter-event time distribution for a
\hpp~with unit rate $\cdf{\tau} = e^{-\tau}$ (black dashed line).  We only
consider inter-event times $\tau>0$ for letter correspondence.
}
\label{fig:rescaled_tau}
\end{figure}

\end{document}